\documentclass{iau}
\usepackage{graphicx}

\title[Chemical Tagging of FGK Stars] 
{Chemical Tagging of FGK Stars: \\  Testing Membership to Young Stellar Kinematics Groups}

\author[D.~Montes et al.]   
{
D.~Montes$^{1}$, 
H.M. Tabernero$^{1,2}$, 
 \and J.I. Gonz\'alez Hern\'andez$^{1,3,4}$}

\affiliation{$^1$Dpto. Astrof\'{\i}sica, Facultad de CC. F\'{\i}sicas, Universidad Complutense de Madrid, E-28040 Madrid, Spain\\email: {\tt dmontes@ucm.es}\\
$^2$Universidad de Alicante, DFISTS, E-03080 Alicante, Spain\\
$^3$Instituto de Astrof\'isica de Canarias, E-38205 La Laguna, Tenerife, Spain\\
$^4$Universidad de La Laguna, Dept. Astrof\'isica, E-38206 La Laguna, Tenerife, Spain}

\pubyear{2015}
\volume{314}  
\pagerange{119--126}
\jname{Young Stars \& Planets Near the Sun}
\editors{J. H. Kastner, B. Stelzer, \& S. A. Metchev, eds.}
\begin{document}

\maketitle

\begin{abstract}
In this contribution talk we summarize the results of our ongoing project of detailed analysis of the chemical content (\textit{chemical tagging}) 
as a promising powerful method to provide clear constraints on the membership of FGK kinematic candidates to stellar  kinematic groups of different ages 
that can be used as an alternative or complementary to the methods that use kinematics, photometry or age indicators.
This membership information is very important to better understand the star formation history in the solar neighborhood discerning between field-like stars (associated with dynamical resonances (bar) or spiral structure) and real physical structures of coeval stars with a common origin (debris of star-forming aggregates in the disk). 
We have already applied the chemical tagging method to constrain the membership of FGK candidate stars to the Hyades supercluster and the Ursa Major moving group and in this contribution we present the preliminary results of our study of the Castor moving group. 
\end{abstract}

\firstsection 
\section{Introduction}

Stellar  kinematic groups (SKGs) --superclusters (SCs) and moving groups (MGs)-- are kinematic coherent groups of stars \cite[(Eggen 1994)]{egg94} that might share a common origin. Among them, the youngest SKGs are: the Hyades SC (600 Myr), the Ursa Major MG (Sirius SC, 300 Myr), the Local Association or Pleiades MG (20 to 150 Myr), the IC 2391 SC (35-55 Myr), the Castor MG (200 Myr), and Hercules-Lyra (see \cite[Montes et al., 2001]{2001MNRAS.328...45M};  \cite{lop06}; \cite{Klutsch14} and references therein). 
More recently other very young SKGs (TW Hya, $\beta$ Pic, Tuc-Hor, AB Dor, Columba and Carina) have been identified with kinematics close to the Local Association  
and others close to other MGs like Argus to IC 2391 and  Octans, Octans-Near to Castor 
(see  \cite{ZS04}; \cite[Torres et al. 2008]{2008hsf2.book..757T}; \cite[Montes 2010, 2015]{mon10, mon15} and references therein). 
Even new associations are identified, such as the ASYA, All Sky Young Association (\cite{ASYA}).
In order to constrain the membership of  kinematic candidate stars to these SKGs additional information is needed. 
Recently, several methods have been developed  to take into account kinematics and photometry simultaneously:
SACY survey (\cite[Torres et al. 2008]{2008hsf2.book..757T}), BANYAN survey (\cite{BANYAN I}), LACEwING (\cite{LACEwING}).
Using different age indicators such as the lithium line at 6707.8 \AA, the chromospheric and coronal activity level, and gyrochronology,
it is possible to quantify the contamination by younger or older field stars (\cite{lop10}; \cite{mal10}).
However, the detailed analysis of the chemical content (\textit{chemical tagging}) is 
another powerful method that provides clear constraints on the membership to these structures 
(\cite{fre02}; \cite{tab12} and references therein). 
Studies of open clusters such as the Hyades and Collinder 261 
have found high levels of chemical homogeneity, showing that chemical information is preserved within
the stars and  that the possible effects of any external sources of pollution are negligible. 
 
\section{Observations and spectroscopic analysis}
Spectroscopic observations were obtained at the 1.2~m Mercator Telescope at the \emph{ Observatorio del Roque de los Muchachos}  (La Palma, Spain) during several observing runs.
Fundamental stellar parameters ($T_{\rm eff}$, $\log{g}$, $\xi$, and [Fe/H]) are determined using an automatic code ({\scshape StePar}) which takes into account the sensitivity of iron $EW$s measured in the spectra, see \cite{tab12}.
Chemical abundances were calculated using the equivalent width ($EW$) method in a fully differential way by comparing them with those derived for a solar spectrum.
A total of 20 elements were analyzed: Fe, the $\alpha$-elements (Mg, Si, Ca, and Ti), the Fe-peak elements (Cr, Mn, Co, and Ni), the odd-Z elements (Na, Al, Sc, and V) and the s-process elements (Cu, Zn, Y, Zr, Ba, Ce and Nd). 
We determined differential abundances $\Delta$[X/H]  (\textit{chemical tagging}) by comparing our measured abundances with those of a reference star known to be a member of the SKG under study on a line-by-line basis.

\begin{figure}[b]
 \vspace*{-0.2 cm}
\begin{center}
\centerline{
\includegraphics[width=2.8in]{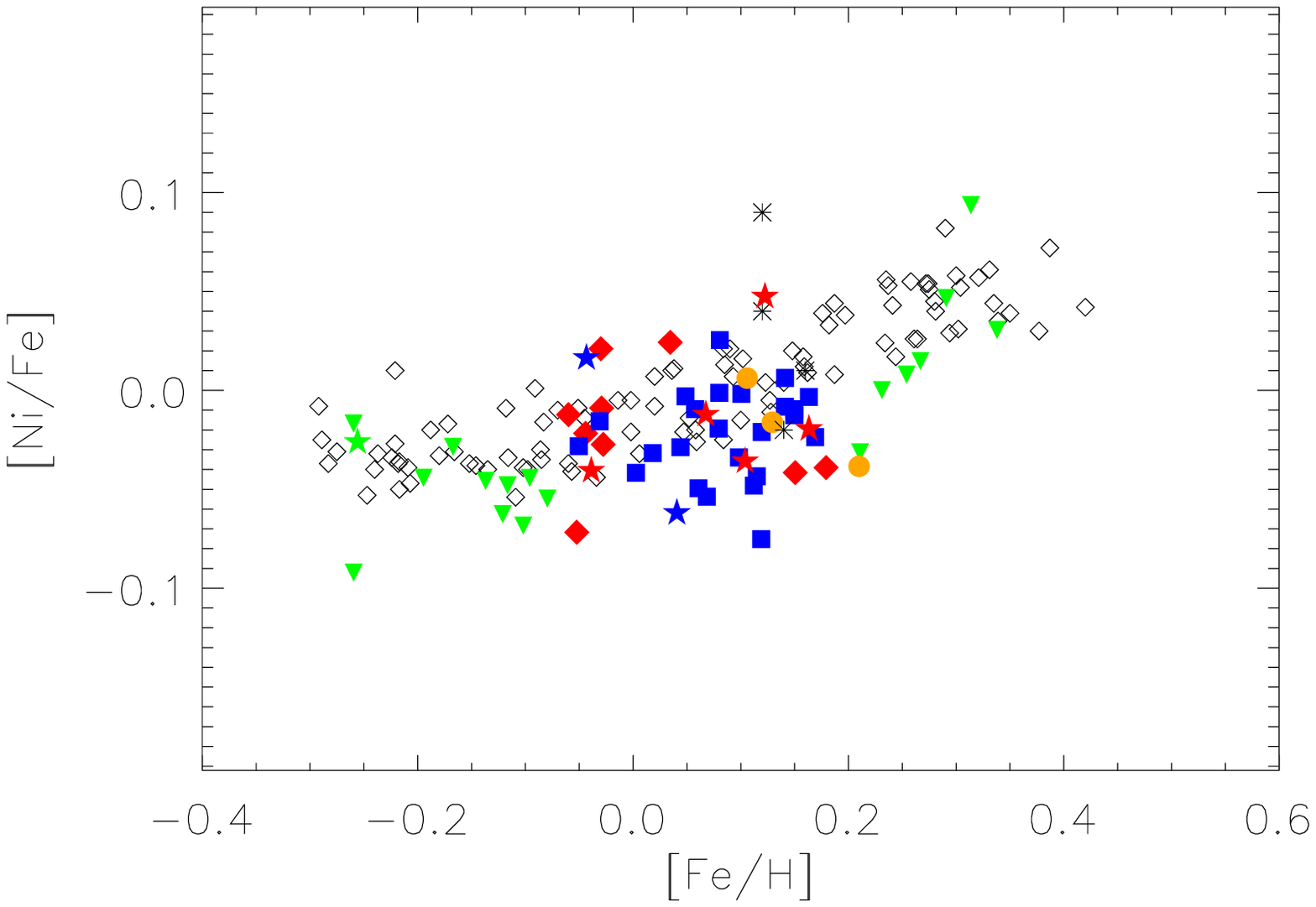}\includegraphics[width=2.8in]{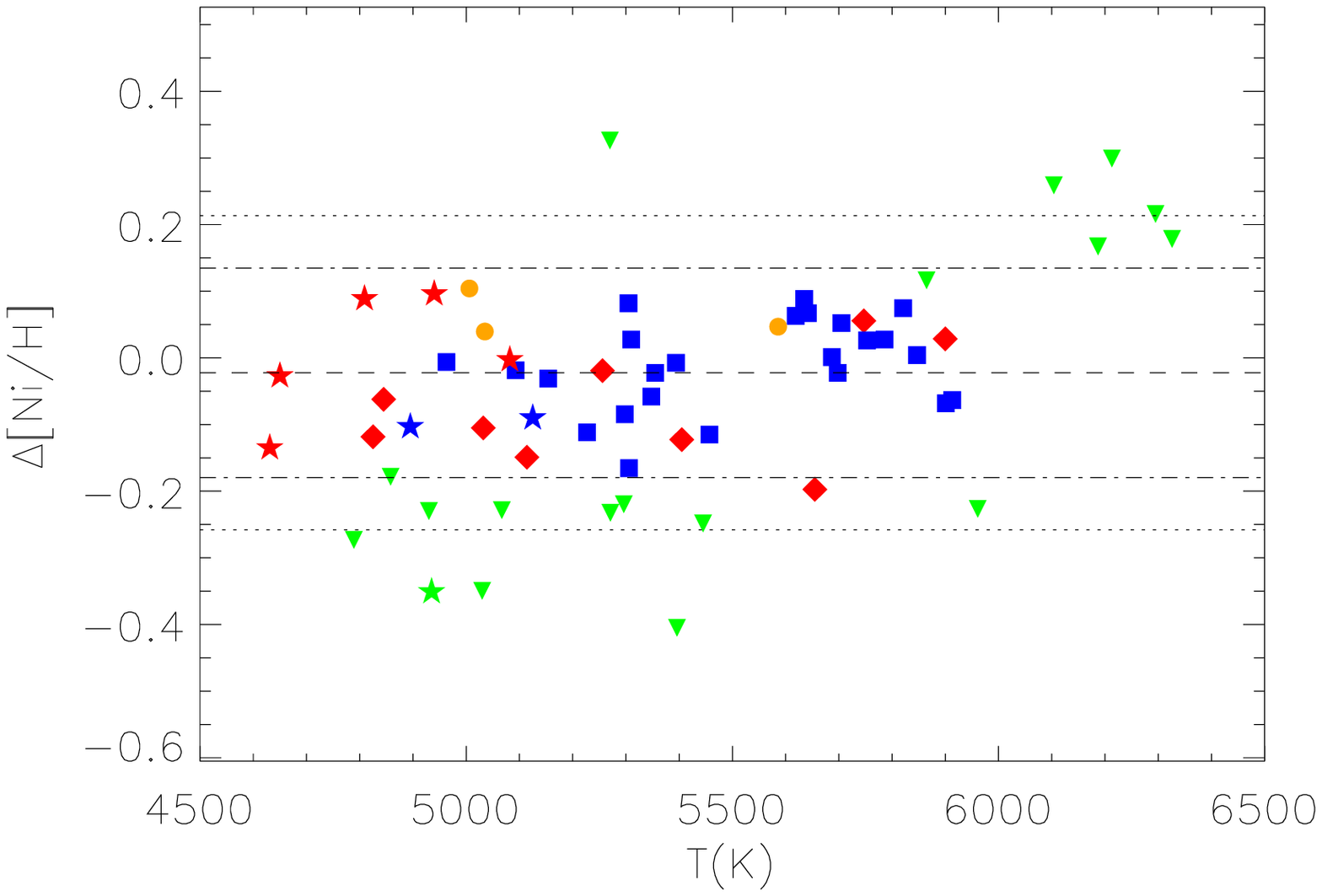}
}
 \vspace*{-0.2 cm}
 \caption{
\textit{Left panel}:  [Ni/Fe] vs. [Fe/H] for the Hyades SC sample, open diamonds represent the thin disc data (\cite[Gonz{\'a}lez Hern{\'a}ndez et al. 2010]{gon10}).
\textit{Right panel}: $\Delta$[Ni/H] (differential abundances for Ni) vs. $T_{\rm eff}$ for the Hyades SC sample. Dashed-dotted lines represent 1-rms over and below the median for our sample,  whereas dotted lines represent the 1.5-rms level. Dashed lines represent the mean differential abundance. 
In both panels, red  diamonds are our stars compatible to within 1-rms with  the Fe abundance but not for all elements, 
blue squares and blue starred symbols are the candidates selected as compatible with Hyades SC membership.
Green downward-pointing triangles show stars incompatible with Hyades SC membership. 
For more details see \cite{tab12}. 
 }
   \label{fig:hya}
\end{center}
\end{figure}

\section{Hyades supercluster}


We have applied the chemical tagging method to constrain the membership of 61 FGK candidate stars to the Hyades SC. 
Fig.~\ref{fig:hya} shows the results for [Ni/Fe] vs. [Fe/H] and $\Delta$[Ni/H] (differential abundances for Ni) vs. $T_{\rm eff}$. For more details see \cite{tab12}. 
We found that 28 of the 61 stars analyzed have homogeneous abundances for all the elements we considered (a 46\% of the studied stars) 
and are compatible with the Hyades isochrone, as expected if they have evaporated from the Hyades cluster. 
The large membership percentage that we find in this work (46 \%) compared with those of other authors 
demonstrates the importance of the sample selection and a detailed chemical analysis.

\section{Ursa Major moving group}

\begin{figure}[h]
 \vspace*{-0.2 cm}
\begin{center}
\centerline{
\includegraphics[width=2.8in]{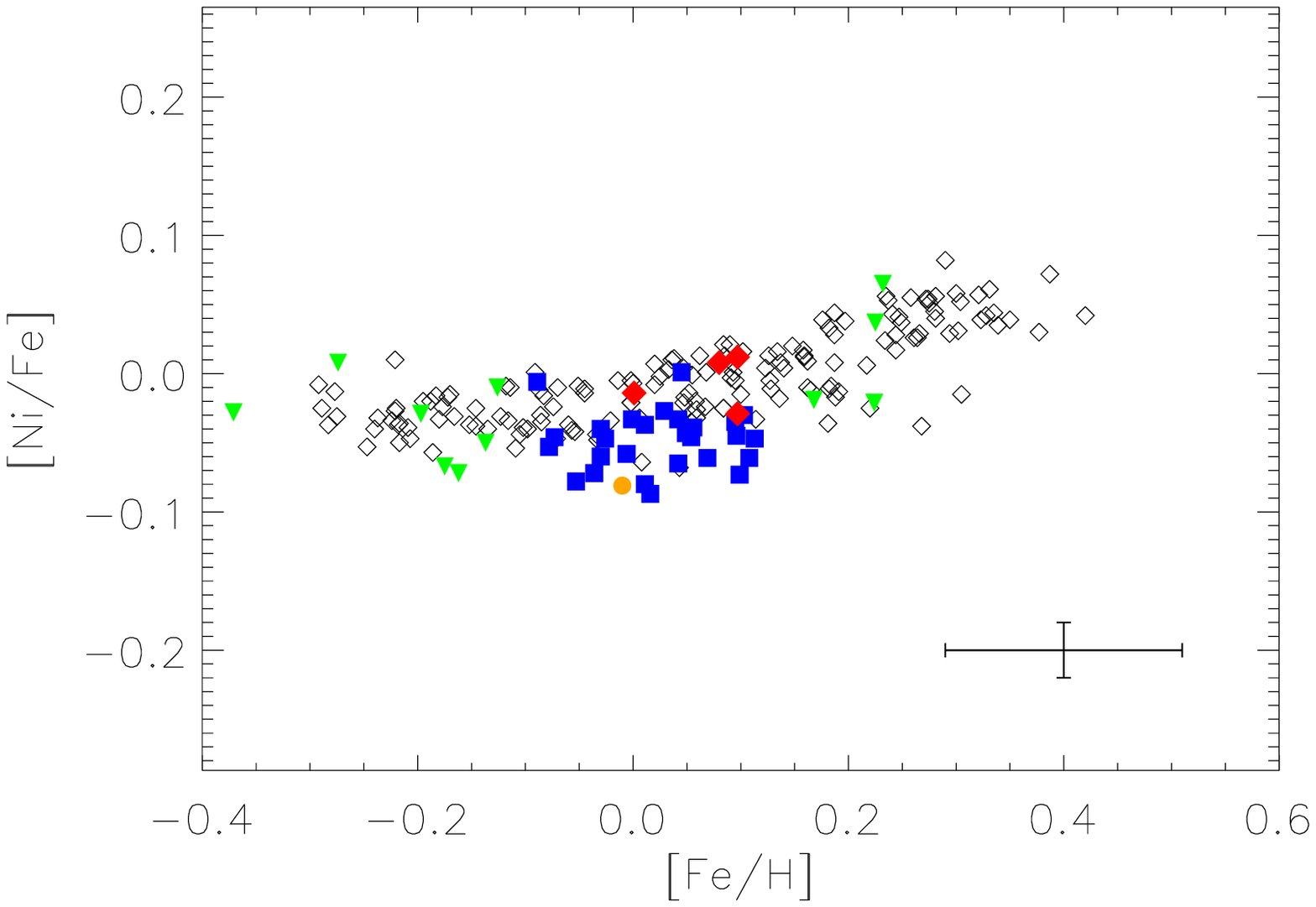}\includegraphics[width=2.8in]{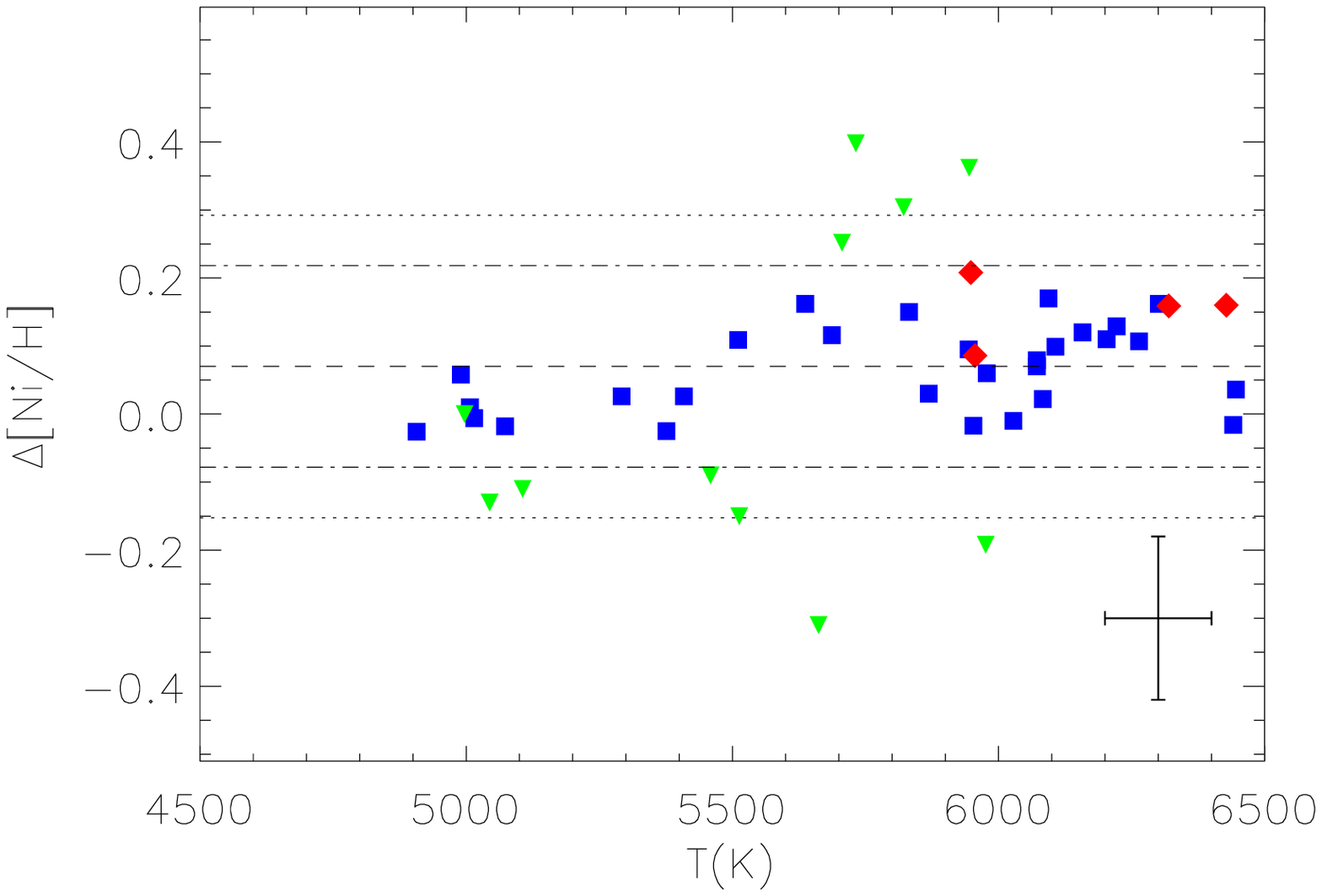}
}
 \vspace*{-0.2 cm}
 \caption{
As Fig.~\ref{fig:hya} for the Ursa Major MG sample, see \cite{tab15}. 
 }
   \label{fig:uma}
\end{center}
\end{figure}

We have applied the chemical tagging method to constrain the membership of 44 FGK candidate stars to the Ursa Major MG.
Fig.~\ref{fig:uma} shows the results for [Ni/Fe] and $\Delta$[Ni/H]  for the Ursa Major MG sample as in Fig.\ref{fig:hya}. For more details see \cite{tab15}. 
Our chemical tagging analysis indicates that the Ursa Major MG is less affected by field 
star contamination than other moving groups (such as the Hyades SC). We find a roughly solar iron composition [Fe/H]=0.03 $\pm$ 0.07 dex for the finally selected stars, whereas the [X/Fe] ratios are roughly sub-solar except for super-solar Barium abundance.
We conclude that 29 out of 44 (i.e. 66 \%) candidate stars share a similar chemical composition. In addition, we find that the abundance pattern of the Ursa Major MG is different from that of the Hyades SC.

\section{Castor moving group}

\begin{figure}[t]
 \vspace*{-0.2 cm}
\begin{center}
\centerline{
\includegraphics[width=2.8in]{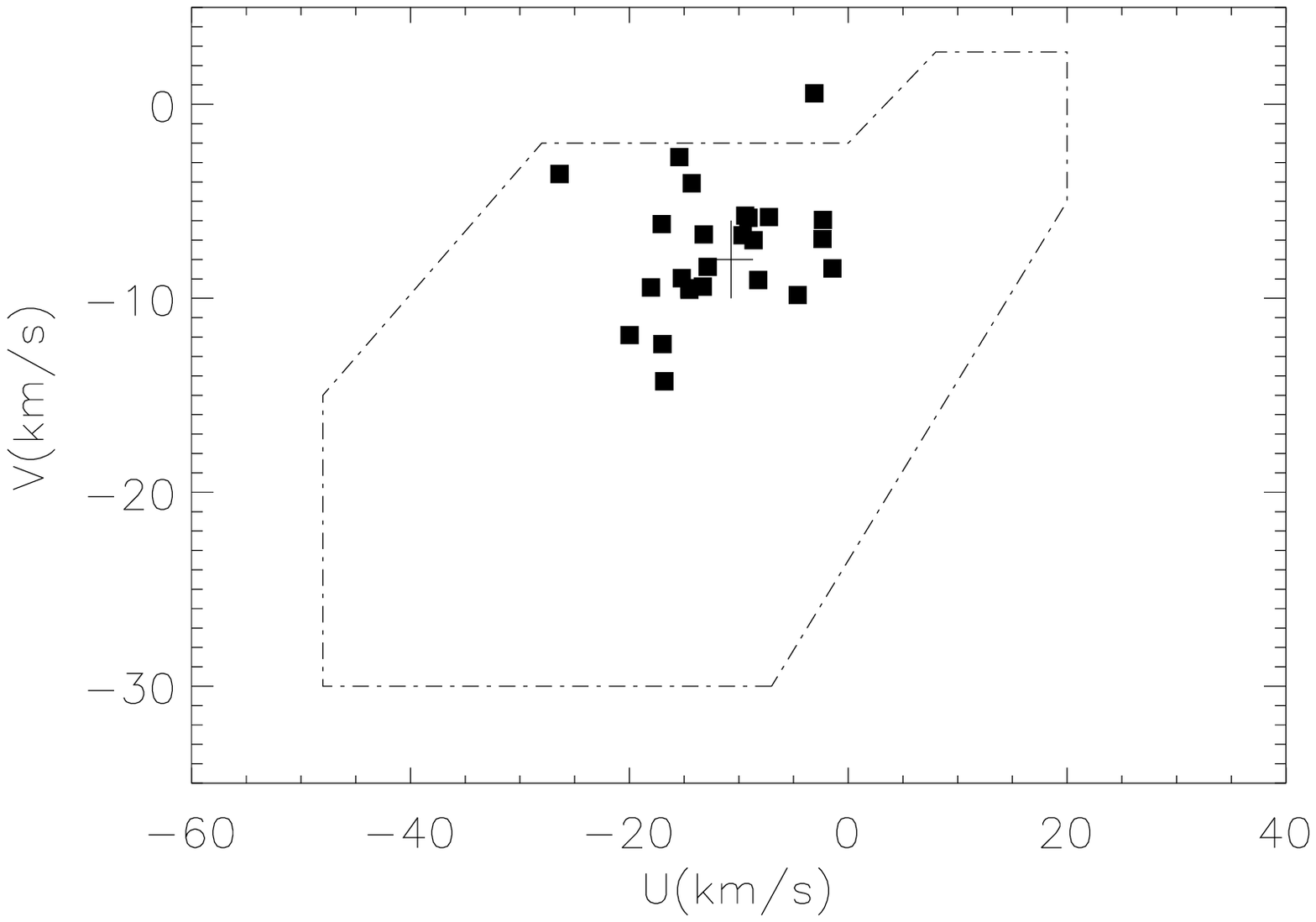}\includegraphics[width=2.8in]{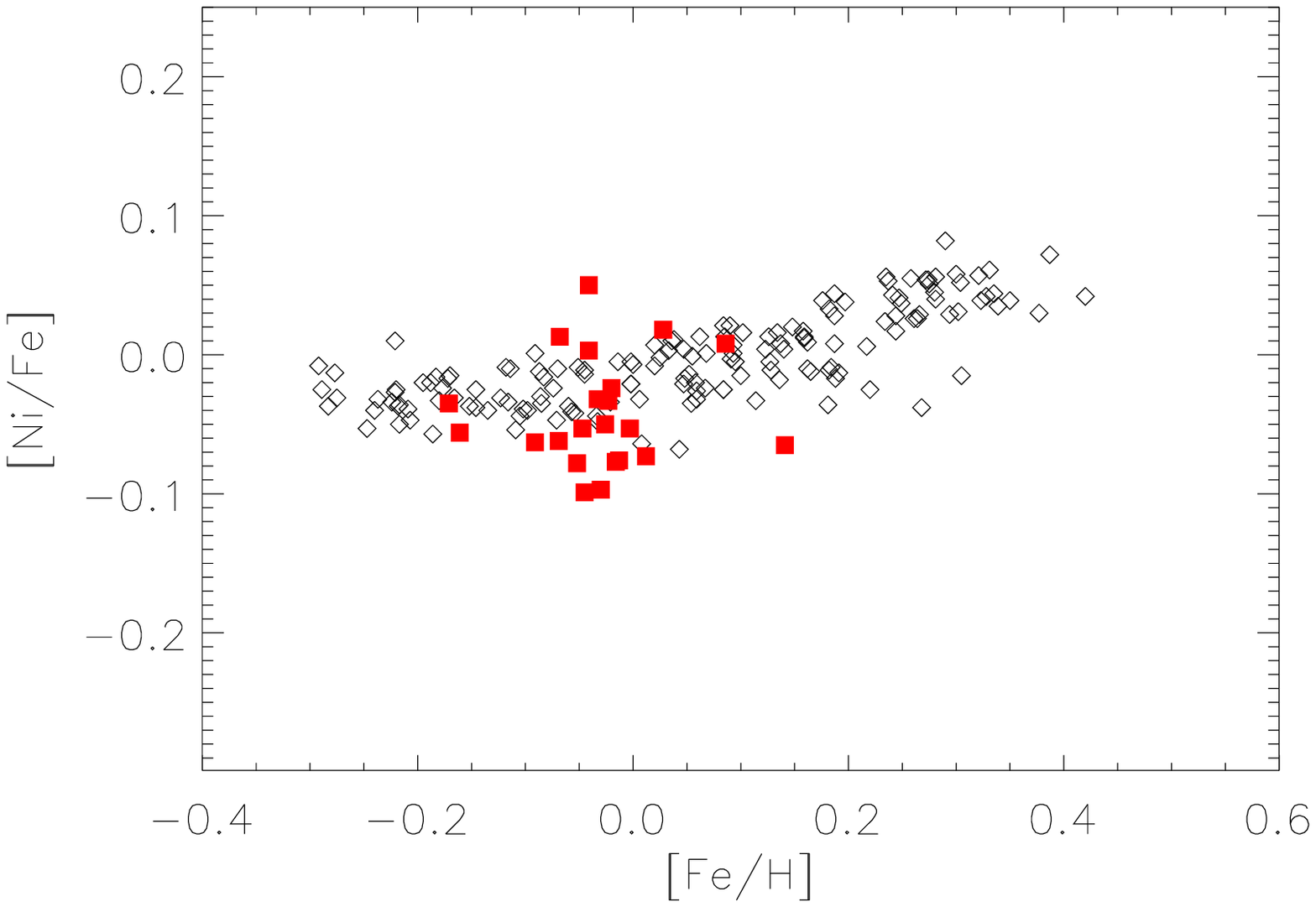}
}
 \vspace*{-0.2 cm}
 \caption{
\textit{Left panel}:  \textit{UV} galactic velocity components for the Castor MG sample.
\textit{Right panel}: [Ni/Fe] vs. [Fe/H] for the Castor MG sample, open diamonds as Fig.~\ref{fig:hya} and Fig.~\ref{fig:uma}
and red  squares represent our sample. For more details see Tabernero et al. (2015, in prep). 
 }
   \label{fig:cas}
\end{center}
\end{figure}

We present  here the preliminary results of our chemical tagging study (Tabernero et al. 2015, in prep) of 19 FGK candidate stars to the Castor MG following the same method used for 
the Hyades SC and the Ursa Major MG (see Fig.~\ref{fig:cas}).
This analysis will contribute to clarify the controversial nature of this MG (see \cite{Mamajek13} and \cite{Mamajek15}) and the possible relation with the Octans and Octans-Near associations.

\section{Conclusions}

We have computed in an homogeneous way the fundamental stellar parameters and differential abundances of FGK candidate stars to different SKGs that allowed us to 
apply the  \textit{chemical tagging} method and constrain their membership and quantify the contamination by younger or older field stars.
We have already applied this method to the Hyades SC, the Ursa Major  and Castor MGs, 
and we plan to apply it to additional spectroscopic observations of other younger SKGs that will contribute to a better understanding of the star formation history in the solar neighborhood discerning between field population stars 
and real physical structures of coeval stars with a common origin. 
%
The \textit{chemical tagging} is therefore a powerful method to confirm known SKGs or identify new ones 
using large high resolution spectroscopic surveys like the Gaia-ESO survey, GES (\cite{Gilmore12}) and GALAH (\cite{GALAH15}).
See also promising approaches (\cite[Mitschang et al. 2013]{mit13}) and limitations (\cite[Blanco-Cuaresma et al. 2015]{BC15})
of the method.

\begin{acknowledgements}
This work was supported by the Univ. Complutense de Madrid (UCM) 
and the Spanish Ministry of Economy and Competitiveness (MINECO) under grants 
AYA2011-30147-C03-02, AYA2011-29060 and Severo Ochoa Program SEV-2011-0187.
\end{acknowledgements}


%


\end{document}